\documentclass[twocolumn,showpacs,preprintnumbers,amsmath,amssymb]{revtex4}

\usepackage{graphicx}
\usepackage{dcolumn}
\usepackage{bm}

\usepackage{url}
\usepackage{epstopdf}

\makeindex
\title{Stabilizing isolated attosecond pulse formation by dispersion tuning}

\begin{document}

\author{Alexander Kothe}
\affiliation{Institut f\"{u}r Experimentalphysik, Universit\"{a}t Hamburg and Center for Free Electron Laser Science (CFEL), Luruper Chaussee 149, 22761 Hamburg, Germany}
\author{Thomas Gaumnitz}
\affiliation{Institut f\"{u}r Experimentalphysik, Universit\"{a}t Hamburg and Center for Free Electron Laser Science (CFEL), Luruper Chaussee 149, 22761 Hamburg, Germany}
\author{Jost Henkel}
\affiliation{Institut f\"{u}r Theoretische Physik, Leibniz Universit\"{a}t Hannover, Appelstra\ss e 2, 30167 Hannover, Germany}
\author{Joachim Schneider}
\affiliation{Institut f\"{u}r Experimentalphysik, Universit\"{a}t Hamburg and Center for Free Electron Laser Science (CFEL), Luruper Chaussee 149, 22761 Hamburg, Germany}
\author{Mark J. Prandolini}
\affiliation{Institut f\"{u}r Experimentalphysik, Universit\"{a}t Hamburg and Center for Free Electron Laser Science (CFEL), Luruper Chaussee 149, 22761 Hamburg, Germany}
\author{Julia Hengster}
\affiliation{Institut f\"{u}r Experimentalphysik, Universit\"{a}t Hamburg and Center for Free Electron Laser Science (CFEL), Luruper Chaussee 149, 22761 Hamburg, Germany}
\author{Markus Drescher}
\affiliation{Institut f\"{u}r Experimentalphysik, Universit\"{a}t Hamburg and Center for Free Electron Laser Science (CFEL), Luruper Chaussee 149, 22761 Hamburg, Germany}
\author{Manfred Lein}
\affiliation{Institut f\"{u}r Theoretische Physik, Leibniz Universit\"{a}t Hannover, Appelstra\ss e 2, 30167 Hannover, Germany}
\author{Thorsten Uphues}
\affiliation{Institut f\"{u}r Experimentalphysik, Universit\"{a}t Hamburg and Center for Free Electron Laser Science (CFEL), Luruper Chaussee 149, 22761 Hamburg, Germany}
\email{Thorsten.Uphues@cfel.de}

\begin{abstract}
The carrier envelope phase (CEP) jitter of isolated attosecond pulses (IAPs) is theoretically and experimentally investigated based on the assumption that a significant contribution originates from the measurable CEP jitter of the driving laser. By solving the time-dependent Schr\"odinger equation, it is demonstrated that the attosecond CEP jitter of IAPs is minimized when the driving pulse is near its Fourier limit but with slightly negative chirp. Although, at present the utilization of the CEP of IAPs has limited applications, understanding and characterization of the CEP jitter of IAPs is the first step towards exact control of the electric field of extreme ultraviolet (XUV) pulses.
\end{abstract}

\maketitle

\section{Introduction}

Stabilization of the carrier-envelope phase (CEP) of femtosecond laser pulses has revolutionized the field of strong-field physics enabling the electronic response of matter to be controlled and measured within one optical cycle \cite{Krausz2014}. In the past 20 years considerable effort has been carried out to actively stabilize the CEP of mode-locked oscillators  \cite{Udem2002} or passively for optical parametric amplifiers \cite{Baltuska2002}. Subsequent chirped-pulse amplification (CPA) of these femtosecond pulses, at a lower repetition rate, requires a separate slow-loop feedback to stabilize the CEP. The remaining CEP jitter of these amplified pulses is dominated by the CPA process through thermal, beam-pointing and mechanical fluctuations \cite{Kakehata2002} and can be measured using various techniques \cite{Wittmann2009,PAULUS2}. An important application of these CEP stabilized femtosecond pulses is the generation of isolated attosecond pulses (IAPs) using high-harmonic generation (HHG) \cite{Krausz2009,Sansone2011}. However, the question arises to which extent the CEP jitter of the driving pulse affects the CEP jitter of the corresponding HHG-based IAP which may present an operational limit to the shortest attosecond pulses demonstrated so far \cite{Li2017,Zhao2012,Gaumnitz2017}.

In a typical attosecond setup using amplitude gating for HHG \cite{Frank2012}, the adjustment of the optical phase and the CEP of the driving laser pulses is usually done by a pair of wedges. Controlling the dispersion by these wedges generates near Fourier-transform-limited driving pulses. Provided the wedges are thin, small changes of the dispersion are treated as CEP control while neglecting the higher order dispersion terms, resulting in a change of the CEP over a range of a few $\pi$ rad.

In this work, we investigate this widely used concept with a complete treatment of the dispersion introduced by the wedges even for small changes in the absolute dispersion. The dependence of the HHG spectrum on the amount of dispersion given by the wedge position, centered around the Fourier-limited pulse, is called a "CEP-scan". In the past, changes in the CEP have been used to identify the half-cycle cutoff (HCO) positions which are extremely sensitive to the CEP and can additionally be used to measure the CEP extremely accurately \cite{Haworth2006}. The characteristic features of HCOs have been further investigated using quantum mechanical calculations by solving the {\em ab initio} time-dependent Schr\"odinger eqation (TDSE) in the single-active electron (SAE) approximation \cite{Henkel2013}.

Furthermore, theoretical studies are carried out to obtain the dependence of the CEP stability of IAPs on the measured CEP jitter of the driving laser pulses. We demonstrate that the CEP-scan technique can be used not only to identify wedge positions for maximum IAP bandwidth corresponding to near Fourier-limited driving laser pulses, but also and more importantly, for minimum CEP jitter of the IAPs. With respect to current developments towards the single-cycle limit of driving laser pulses \cite{Goulielmakis2008a,Wirth2011b,Mucke2015a,Manzoni2015}, we believe that adequate control of the attosecond pulse is required and we demonstrate a first approach to stabilize the attosecond pulse formation by dispersion engineering for exremely short driving laser pulses.


\section{CEP-scan}\label{sec:CEPscan}

\begin{figure*}[ht!]
\begin{center}
\includegraphics[clip=true,trim=0cm 0cm 0cm 0cm,width=0.95\linewidth]{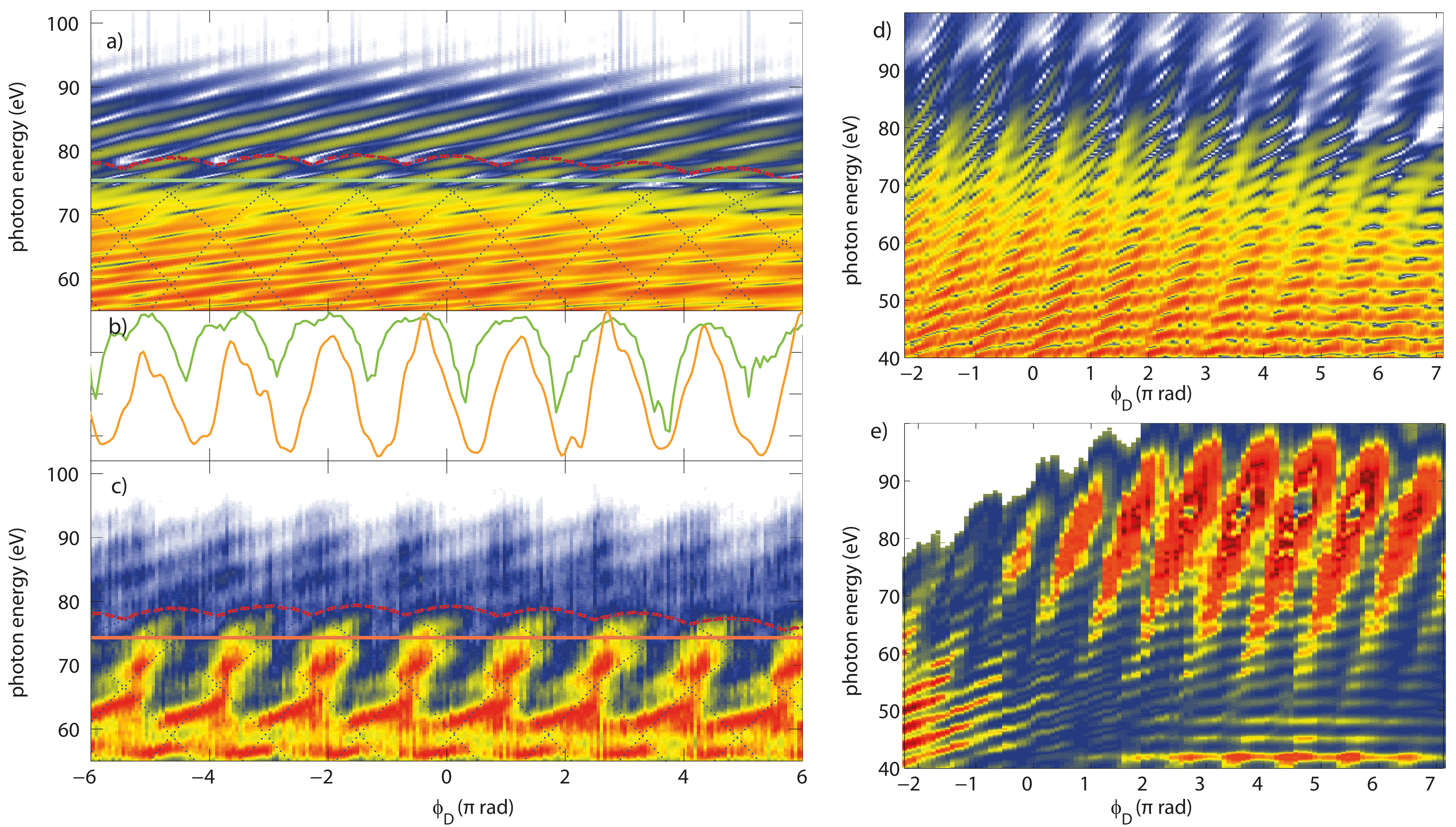}%
\end{center}
\caption{Left panel: Comparison of the CEP-scans obtained through TDSE calculations (a) and in the experiment (c), where calculations are based on the spectral and temporal shape for the $5.9$~fs pulse measured in the experiment (intensity of $4\cdot10^{14}$~W/cm$^2$). The cut-off energy calculated with the intensity model is given by the red dashed lines. The dotted blue lines are the half-cycle cut-offs. The intensity profiles at $E=79$~eV photon energy from both the TDSE calculations (solid green line) and experiment (solid orange line) are compared in (b) (adapted from \cite{Gaumnitz2015a}).  Right panel: a CEP-scan comparison between TDSE calculations (d) and experiment (e) (adapted from Frank {\em et al.} \cite{Frank2012}) for a $4$~fs pulse.}
\label{fig:JostMeasurement-SEAF}
\end{figure*}

We use a CEP-stabilized sub-6~fs few-cycle Ti:Sa-laser system to generate isolated attosecond pulses by amplitude gating and spectral filtering of the HHG cut-off region \cite{Gaumnitz2015a}. The CEP defines the form and the length of the continuum region at the cut-off \cite{Constant1999}. Experimentally, the CEP is controlled by a pair of wedges introducing either a positive or negative chirp to a Fourier-limited pulse by material dispersion. The CEP stabilization and control are realised with a f-to-0f interferometer in the oscillator and a f-to-2f interferometer in the amplifier averaging over 30 shots. For this laser system, the CEP jitter was measured to be $60$~mrad. A neon-gas filled hollow-core fiber spectrally broadens the pulses and a set of chirped mirrors compresses the octave spanning spectrum to Fourier-transform-limited pulses of 5.9~fs. For  HHG, the beam is focused into a neon filled target with a propagation length of $\approx\!2$~mm. HHG spectra were recorded over many CEP periods as shown in Fig.~\ref{fig:JostMeasurement-SEAF}c), with a slit grating spectrometer to identify the highest cut-off energy in the HHG spectrum.

\subsection{CEP-Scan: Simulations}
Classical and quantum mechanical simulations including the full dispersion of the wedges (CEP-scans) were performed and compared with the experimental results. First, the HHG process is modeled classically using the intensity model (see \cite{Gaumnitz2015a}), where the linear polarized laser electric field is expressed as:
\begin{equation}
E(t)=E_L(t)\cos{(\omega_0 t+\phi_D)}\label{Eq:PulseTime}\text{,}
\end{equation}
where $E_L(t)$ is the time-dependent field envelope and $\omega_0$ is the carrier frequency. The CEP of the driving laser is described by $\phi_D$. For few-cycle pulses, the electric field extrema vary significantly in amplitude for adjacent half-cycles, leading to different cut-off energies in the HHG process. Interference of the half-cycle trajectories leads to a modulation of the spectrum. A continuum is generated, where no interference occurs. The classical cut-off depends linearly on the ionization potential of the medium ($I_p$) and the instantaneous intensity ($I$): $E_{cut}=3.17U_p+I_p$, so $E_{cut} - I_p \propto\lambda^2I$ \label{eq:Ecut} \cite{Corkum1993}. Therefore, the following simple relation for the energy range of the continuum is given as
\begin{equation}
	E_{continuum}=(E_{cut}-I_p)\frac{\mid I_{max1}-I_{max2}\mid}{I_{max2}}\text{,}
\end{equation}
where $I_{max1}$ and $I_{max2}$ are the values of the intensity maxima of the two neighboring half-cycles contributing to the highest energy part of the HHG process. The CEP dependency leads to values between zero for sine-like pulses and a maximum for cosine-like pulses. Second, the classical HCO model is a similar approach, but expands to all local field extrema \cite{Haworth2006}. With this model, the impact of the individual half cycles of a few-cycle pulse leads to localized features in the harmonic spectrum connected to the individual half-cycles of the driving pulse.

Third, for the quantum mechanical simulations, the electron dynamics leading to HHG is described by using the time-dependent Schrödinger equation (TDSE). TDSE calculations in single-active electron (SAE) approximation with intensity averaging were carried out, revealing a strong impact of the spectral phase on the HHG cut-off behavior \cite{Henkel2013}. A deviation from the spectral phase of a Fourier-transform-limited pulse leads to a reduced cut-off energy and shorter continuum.

For the simulations, the spectral and temporal shape of the initial pulse $E_{in} (t)$ for $\phi_D = 0$~rad is taken from the experiment measured with a SEA-F-SPIDER \cite{Kosik2005} and is Fourier transformed into the frequency domain together with the dispersion introduced by the wedges. In the following we distinguish between a pure CEP shift denoted as $\phi_{CEP}$ and a CEP shift introduced by dispersion (CEP-scan) denoted as $\phi_D$. Thus, the electric field $\tilde{E}(\omega)=\mathcal{FT}\{E_{in}(t)\}$ is multiplied with an additional phase term according to $\phi_D$ including the higher order dispersion. Spectrum and phase are then transformed back to the time domain. The resulting pulse shape used for further calculations is given by:
\begin{equation}
	E_{out}(t)=\mathcal{FT}^{-1}\left\{\mathcal{FT}\{E_{in}(t)\}e^{-i\frac{\omega}{c}\tilde{n}(\omega)L}\right\}.
	\label{eq:EfieldChirp}
\end{equation}
The fused-silica index of refraction is taken from Kitamura {\em et al.} \cite{Kitamura2007}. According to these values a change of the propagation distance in glass of $L=28$ $\mu$m results in a change of $\phi_{CEP}$ by $\pi/2$\,rad at a carrier wavelength of $\lambda_c= 800$~nm.

We performed the TDSE simulations in the single-active electron approximation in a two-dimensional (2D) model for the neon atom. Absorbing boundary conditions similar to \cite{Henkel2011} were employed to avoid unphysical reflections or transmissions at the grid borders. Further details for the TDSE simulation are discussed in \cite{Henkel2013}.

\subsection{CEP-scan: Comparison of experiment and simulation results}

The comparison between two experimental CEP-scans and TDSE simulations is shown in Fig.~\ref{fig:JostMeasurement-SEAF}.
In Fig.~\ref{fig:JostMeasurement-SEAF}c), a CEP-scan from the laser described in section~\ref{sec:CEPscan} is given with a pulse duration of 5.9~fs pulse and an IR intensity of $4\cdot 10^{14}$~W/cm$^2$ in neon. The spectral region below 75~eV is assigned to the HHG-plateau region. The region above 65~eV is strongly modulated when $\phi_{D}$ is varied. The less intense part above 75~eV is assigned to be the cut-off region, which is used in experiments to generate the IAP. The cut-off is reproduced with the intensity model indicated by the red dashed line. Fig.~\ref{fig:JostMeasurement-SEAF}a) presents the corresponding results from the TDSE calculations using the parameters from the experiment. Experiment and simulations reveal spectrally shifting HHG lines above the classical cut-off, which cannot be attributed to a propagation effect as the TDSE calculations do not include such effects.
Line-outs at 79~eV (green and orange solid lines in Fig.~\ref{fig:JostMeasurement-SEAF}a) and c)) compare the simulation to experiment.
Without dispersion, the periodicity is expected to be $n\cdot \pi$~rad. However, in a CEP-scan the dispersion is varied and therefore the resulting electric field is not strictly periodic.

Good agreement with our theory is also obtained for a CEP-scan taken from Frank \textit{et al.} \cite{Frank2012}. In Fig.~\ref{fig:JostMeasurement-SEAF}~d) and e), the simulated spectra and the energy filtered experimental spectra for a CEP scan are shown, respectively. Note, that the simulations consider a 4 fs pulse with Gaussian temporal and spectral shape at an intensity of $4\cdot 10^{14}$~W/cm$^2$. In the energy range from 40 to 60 eV of both the experimental and the simulated spectra clear diagonal ($\phi_D \lesssim 2$) and horizontal ($\phi_D \gtrsim 2$) structures are visible, which can be used to determine $\phi_D=0$~rad in the experiment. With these two comparisons to experimental data, we rely on the simulations to extract an absolute $\phi_D=0$~rad value for calibration.

\section{Stabilizing the IAP generation}

\begin{figure}[t!]
\begin{center}
\includegraphics[clip=true,trim=0cm 0cm 0cm 0cm,width=0.8\linewidth]{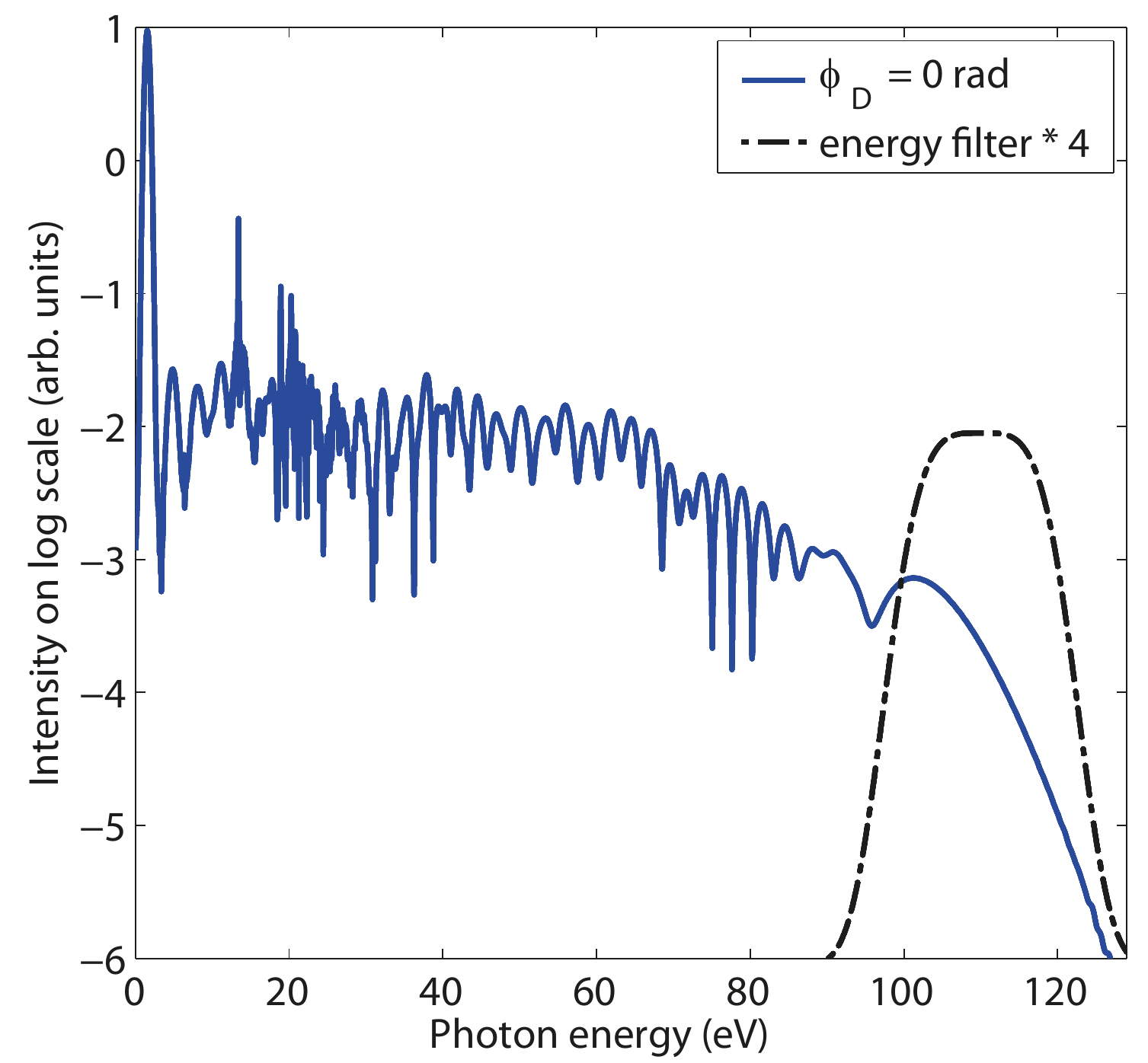}%
\end{center}
\caption{HHG spectrum for a 3.8~fs pulse at an intensity of $I=4\cdot10^{14}$\,W/cm$^2$ for $\phi_D=0$ (blue solid line) and the applied spectral filter (black dash-dotted line), (adapted from \cite{Gaumnitz2015a}). \label{fig:AttoBurst}}
\end{figure}

\begin{figure}[b!]
\includegraphics[clip=true,trim=0cm 0cm 0cm 0cm,width=\linewidth]{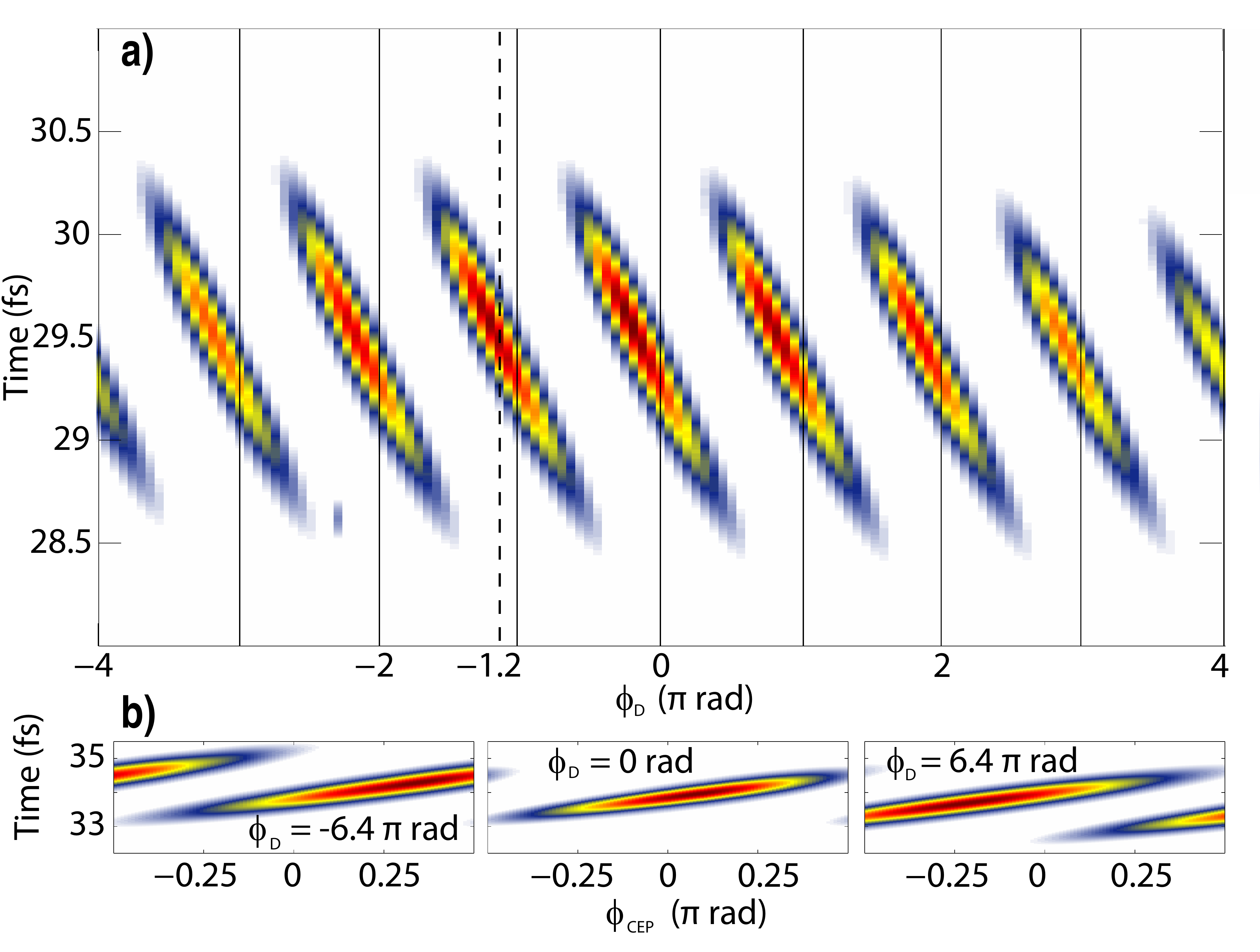}%
\caption{a) The resulting Fourier transformed CEP-scan IAPs (in the time domain) calculated on the basis of a 3.8~fs Fourier-limited pulse duration at $\phi_D=0$ by applying the supergaussian filters at $E_{center}=110$~eV as shown in Fig. \ref{fig:AttoBurst}. In this case, CEP phase change $\phi_D$ is introduced by glass wedges causing additional optical pulse dispersion. The calculated HHG pulses are shown in a false color representation. The dashed vertical line represents the $\phi_{D}$ value with the smallest IAP CEP jitter. b) Dispersion-free scan (for varying $\phi_{CEP}$) for three chirped pulses at $\phi_D = -6.4$~rad (left), $\phi_D = 0$ (center), and $\phi_D = 6.4$~rad (right), (adapted from \cite{Gaumnitz2015a}). \label{fig:AttoPulseFormation38fs}}
\end{figure}

In order to demonstrate optimal IAP generation, further calculations are conducted with ideal pulses with a Fourier-limited duration of 3.8~fs and intensity of $4\cdot10^{14}$~W/cm$^2$. As an example, a single HHG spectrum for such a Fourier-limited driving pulse at $\phi_D = 0$ is shown in Fig.~\ref{fig:AttoBurst}. The black dash-dotted line indicates the spectral filter that is applied to generate IAPs. The corresponding temporal shape is given in Fig.~\ref{fig:AttoPulseFormation38fs}a) for the complete CEP-scan, depicting the attosecond bursts.

We investigate the influence of the CEP jitter on the HHG process by TDSE calculations for 3 positions in the CEP-scan at $\phi_D =-6.4 \pi$~rad (negative chirp), $\phi_D =0$~rad, and $\phi_D = 6.4 \pi$~rad (positive chirp) and varying $\phi_{CEP}$ in the range $\pm\!100$~mrad to mimic the jitter of the laser system without dispersion (see Fig.~\ref{fig:AttoPulseFormation38fs}b)).
Note, we assume no further dispersive effects included in $\phi_{CEP}$. For the negatively and the positively chirped pulses, CEP jitter can lead to the formation of two attosecond pulses with a probability of $\approx\!50$\%. Only in the Fourier limited case ($\phi_D =0$, see Fig.~\ref{fig:AttoPulseFormation38fs}b) (center)) the additional jitter of the CEP allows for a single IAP almost over the full range of $\phi_{CEP}$.

\begin{figure}
\includegraphics[clip=true,trim=0cm 0cm 0cm 0cm,width=0.48\textwidth]{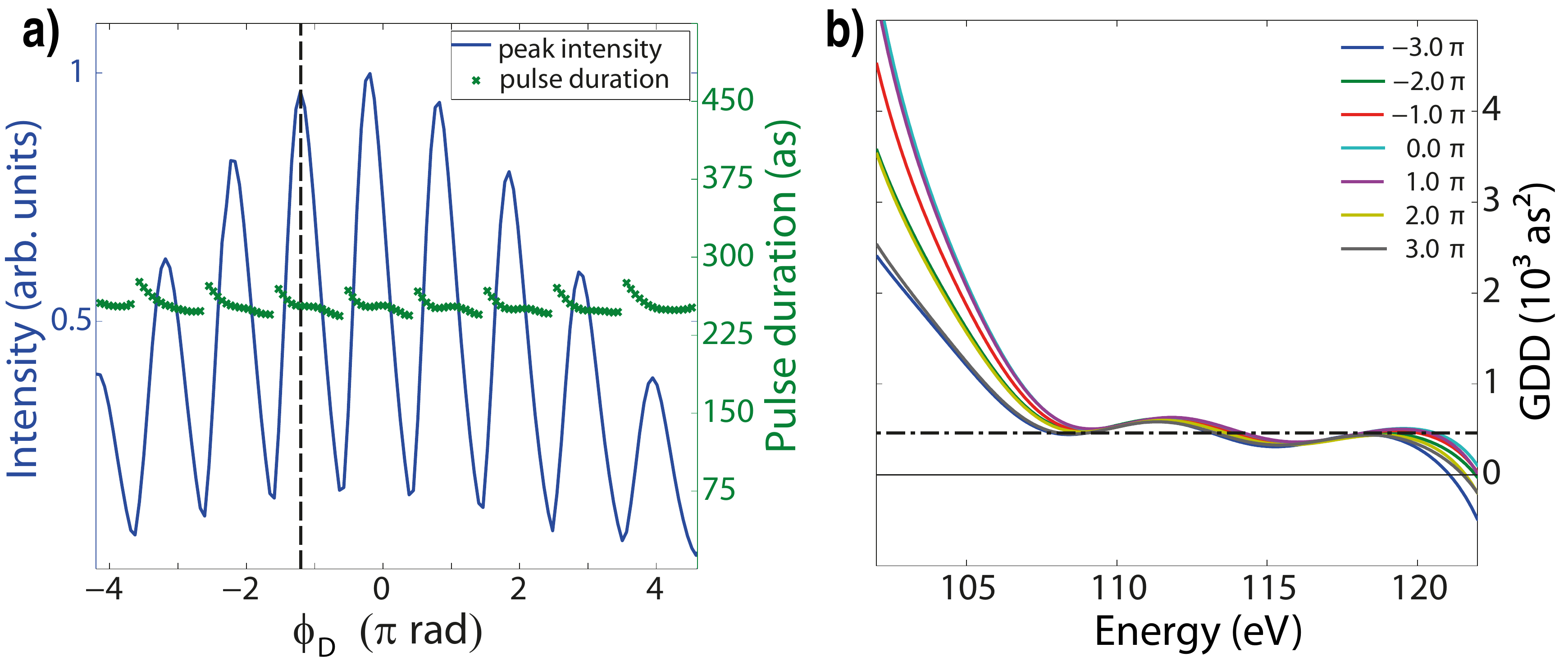}%
\caption{a) Variation of peak intensity (blue line) versus $\phi_{D}$. In addition, the IAP durations are plotted in green. The dashed vertical line represents the $\phi_{D}$ value with the smallest IAP CEP jitter (see text). b) Spectral phase of IAPs generated at positions $\phi_{D}=-3, -2,...,3\pi$\,rad, (adapted from \cite{Gaumnitz2015a}). \label{fig:AttoChirp}}
\end{figure}

To determine the optimal IAP, we need to consider the CEP-scan phase $\phi_D$ with the largest IAP peak intensity. The peak intensity is shown in Fig.~\ref{fig:AttoChirp}~a) by the blue line, while the green crosses indicate the pulse durations for the different values of $\phi_D$. The rising edges in the intensity graph lead to longer IAP durations compared to the duration at maximum intensities. The shortest pulse durations can be achieved at the falling edge of the intensity slopes, but with the drawback of smaller IAP peak intensity. In the regions close to the intensity maxima, the pulse durations stay constant.

Next, we consider the spectral phase (GDD) of the IAPs at distinct positions of $\phi_D$ from $-3.0\pi$~rad to $3.0\pi$~rad shown in Fig.~\ref{fig:AttoChirp}b). Here, only slight differences in phase (GDD) are observed around the central photon energy of the attosecond pulse between 110~eV and 115~eV. At lower and higher photon energies the differences are more clearly visible. These differences lead to different temporal pulse shapes, and therefore, slightly different IAP pulse durations.

\begin{figure}[t!]
\includegraphics[clip=true,trim=0cm 0cm 0cm 0cm,width=\linewidth]{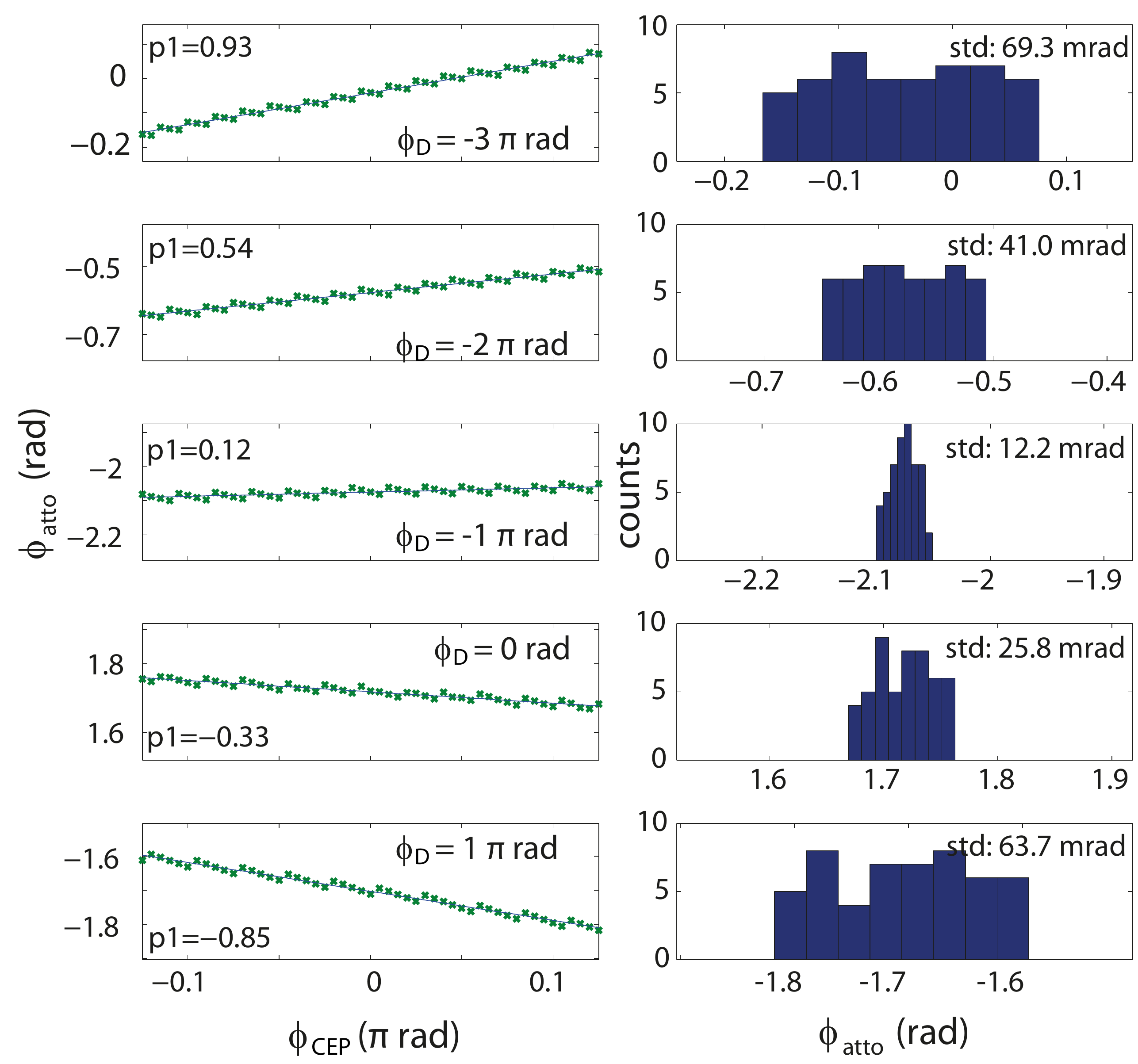}%
\caption{Left panel: The dependence of the CEP of the IAP ($\phi_{atto}$) on the CEP of the driving laser ($\phi_{CEP}$) is calculated for five driving laser pulses at $\phi_{D}=-3\pi, -2\pi,...,1\pi$~rad with different chirp. The dependence of $\phi_{atto}$ on $\phi_{CEP}$ is approximately linear over the considered range, where $p1$ is the slope. Right panel: Corresponding histograms from the results of the left panel, assuming a driving laser CEP jitter from $-100$ to $100$~mrads. For each case, the corresponding $\phi_{atto}$-jitter (std.) in rms is given, (adapted from \cite{Gaumnitz2015a}).  \label{fig:JitterSlope}}
\end{figure}

The influence of the CEP-jitter of the driving laser on the CEP of the IAP ($\phi_{atto}$) is depicted in Fig.~\ref{fig:JitterSlope}. Five positions from a CEP-scan of the driving laser at $\phi_D = -3 \pi$~rad to $1 \pi$~rad are considered. For each $\phi_D$, the dependence of $\phi_{atto}$ on driving laser CEP without dispersion is calculated (Fig.~\ref{fig:JitterSlope} (left panel)). In each case, the dependency is found to be approximately linear over the considered $\phi_{CEP}$ range; the slopes of the linear dependency are provided in Fig.~\ref{fig:JitterSlope} (left panel). For the strongest up- and down-chirp, the slopes are steeper than the slope at $\phi_D = -1\pi$~rad. Corresponding histograms are generated (Fig.~\ref{fig:JitterSlope} (right panel)) for the CEP jitter of the IAP ($\phi_{atto}$), assuming a CEP-jitter of the driving laser from $-100$ to 100~mrads. Clearly in this case, the choice for maximum IAP intensity at $\phi_D$ just below 0 (see Fig.~\ref{fig:AttoChirp}~a)) does not achieve the minimum possible CEP jitter of the IAP at $\phi_D = -1.2\pi$~rad (see dashed line in Fig.~\ref{fig:AttoPulseFormation38fs}~a) and \ref{fig:AttoChirp}~a)). Thus, the CEP jitter of the IAP, $\Delta \phi_{atto}$, is smallest for slightly negative chirped pulses.

\section{Discussion}
We demonstrate the possibility to optimize the driving laser's spectral phase $\phi_D$ to generate ideally stable attosecond pulses using TDSE simulations including dispersion-free jitter $\phi_{CEP}$. We found CEP-scan regimes of the driving laser pulse, where an additional jitter may result in the generation of two attosecond pulses of low intensity. Generally, by changing $\phi_D$ in a CEP-scan, a value for a theoretical minimum of the CEP jitter of the attosecond pulses $\phi_{atto}$ can be identified. This minimum does not necessarily coincide with the most intense attosecond pulse and therefore a compromise between ideal CEP stability and intensity must be made. This value is accessible in the experiment by choosing the corresponding dispersion controlled by the pair of wedges. According to our simulations, a slight negative chirp of -1.2$\pi$~rad yields optimal IAP formation and stability.

\bibliography{BiBAttoHHG}

\section*{Acknowledgements}
We thank J. W. G. Tisch for discussions and support for the HCO calculations as well as the kind permission to reproduce Fig. 22 from \cite{Frank2012} for our analysis.\\
We are grateful for financial support of the Landesexzellenzcluster "Frontiers in Quantum Photon Science". T.G. and J.H.$^1$ acknowledge funding from the Joachim Herz foundation. J.H.$^1$ acknowledges support from the IMPRS UFAST. J.H.$^2$ acknowledges funding from the EU Marie Curie Initial Training Network FASTQUAST. We acknowledge financial support from the Deutsche Forschungsgemeinschaft.

\section*{Additional information}

We declare that we have no significant competing financial, professional or personal
interests that might have influenced the performance or presentation of the work
described in this manuscript.

\end{document}